\documentstyle[aps,prb,preprint]{revtex}
\begin{document}
\draft

\title{Phase Transitions in a Model Anisotropic High $T_{\rm c}$ 
Superconductor}

\author{Tao Chen and S. Teitel}

\address{Department of Physics and Astronomy, University of Rochester,
Rochester, New York 14627}

\date{\today}

\maketitle

\begin{abstract}
We carry out simulations of the anisotropic
uniformly frustrated 3D XY model, as a model
for vortex line fluctuations in high $T_{\rm c}$ superconductors.  
We compute the phase diagram as a function of temperature and
anisotropy, for a fixed applied magnetic field $B$.  
We find that superconducting coherence parallel to $B$ persists
into the vortex line liquid state, vanishing at a $T_{{\rm c}z}$ 
above the melting $T_{\rm m}$.  Both $T_{{\rm c}z}$ and $T_{\rm m}$ are
found in general to lie well below the cross-over $T_{\rm c2}$
from the vortex line liquid to the normal state.
\end{abstract}


\section{Introduction}

From a phenomenological point of view, ``high $T_{\rm c}$''
superconductors are believed to differ from conventional  
type II superconductors primarily because of 
the dramatically enhanced importance of thermal 
fluctuations.\cite{R1,R2}
In an applied magnetic field ${\bf H}$, such thermal
fluctuations are believed to melt the Abrikosov vortex line lattice
at a temperature $T_{\rm m}$ well below the mean field $T_{\rm c2}$ which marks
the onset of strong diamagnetism.\cite{R3,R4}  
In between $T_{\rm m}$ and $T_{\rm c2}$ is
a new vortex line liquid state.  
Experimentally, this picture has been supported by the
observation that, in high $T_{\rm c}$ materials, the onset of reversible 
diamagnetism occurs at
a temperature well above that where resistance vanishes;\cite{R5}
the separation between these temperatures increases
with increasing $H$.  
According to this picture, the onset of diamagnetism
at $T_{\rm c2}$ is
associated with a growth in local superconducting correlations, giving rise
on short length scales to a finite superconducting
wavefunction $\psi({\bf r})$ in terms of which vortex lines can be defined.
This $T_{\rm c2}$ marks a strong cross-over region, rather than a sharp
thermodynamic transition.
In the resulting vortex line liquid, free diffusion of vortex lines gives
rise to ``flux flow'' electrical resistance. 
The vanishing of resistance
only occurs at a lower temperature when the line liquid freezes into
a lattice or glass.  

To investigate the effect of thermal fluctuations on phase transitions
in type II superconductors, within a numerical simulation,
one of us and Y.-H. Li have previously introduced\cite{R6,R7} the
three dimensional (3D) uniformly frustrated XY model.
Simulations of this model in the isotropic coupling limit, at low
vortex line density, gave the surprising result that superconducting
coherence parallel to the applied magnetic field appeared to persist
above the vortex line lattice melting temperature, into the vortex 
line liquid phase.\cite{R7}  The goal of the present work is to extend these
simulations to a model with  uniaxial anisotropic couplings, so as
to better model the layered structure of the high $T_{\rm c}$ materials.
We will consider only the case where the applied magnetic field is
parallel to the anisotropy axis ${\bf\hat z}$.  We will map out the
phase diagram in the anisotropy--temperature plane, looking for the
presence of parallel coherence in the vortex liquid phase and 
dimensional cross-over as anisotropy increases.  
The rest of this paper is organized as
follows.  In Section \ref{smodel} we describe our model, its
limits of validity, and the specific parameters of our simulations.
In Section \ref{sresults} we give our numerical results, mapping out
the phase diagram, and characterizing the nature of 
vortex line fluctuations.  In Section \ref{sdisc} we discuss our results
and present our conclusions.

\section{Model}
\label{smodel}

Our model  
arises from the familiar Ginzburg-Landau (GL) free energy functional 
with the following additional approximations: ($i$) the continuum is 
discretized to a periodic grid of sites, ($ii$) the amplitude of the
superconducting wavefunction is taken to be constant outside of
a vortex core (London approximation), and ($iii$) the internal
magnetic field ${\bf B}$ is taken to be a spatially uniform fixed
constant.  These yield the Hamiltonian,\cite{R7}
\begin{equation}
  {\cal H}[\theta_i]=-\sum_{\langle ij\rangle}J_{ij}\cos
  (\theta_i-\theta_j-A_{ij}) \enspace ,
\label{eHjj}
\end{equation}
where $\theta_i$ is the phase angle of the wavefunction at site $i$ of the 
discrete grid, the sum is over nearest neighbor bonds $\langle ij\rangle$, 
$A_{ij}=(2\pi/\phi_0)\int_i^j{\bf A}\cdot d{\bf\ell}$
is the integral of the magnetic vector potential across the bond
($\phi_0=hc/2e$ is the flux quantum),
and $J_{ij}=J_z$ or $J_\perp$ is the superconducting coupling in
the direction of the bond.
We consider here a cubic grid of sites. 
If we identify the discrete spacing along ${\bf\hat z}$
as the distance $d$ between copper-oxide planes, and
the discrete spacing in the $xy$ plane as representing the short
length cutoff for a vortex core, given by the coherence
length $\xi_\perp$, we have
\begin{equation}
   J_\perp\equiv {\phi_0^2d\over 16\pi^3\lambda_\perp^2}\enspace ,\qquad
   J_z={\phi_0^2\xi_\perp^2\over 16\pi^3\lambda_z^2d}\enspace ,
\label{eJ}
\end{equation}
where $\lambda_\perp$ and $\lambda_z$ are the magnetic penetration
lengths within and normal to the copper-oxide planes, respectively.  We
define an anisotropy parameter $\eta$ as
\begin{equation}
  \eta\equiv\sqrt{J_\perp\over J_z}={\lambda_z\over\lambda_\perp}
  {d\over\xi_\perp} \enspace .
\label{eta}
\end{equation}
Note that if the coherence length along ${\bf\hat z}$ is larger
than the interplanar spacing, $\xi_z>d$, then one should replace
$d$ by $\xi_z$ in Eqs.(\ref{eJ}-\ref{eta}) above.  In this case,
since the GL free energy with anisotropic masses gives\cite{R1} 
$\xi_z/\xi_\perp=\lambda_\perp/\lambda_z$, we have
$\eta=1$ and hence isotropic couplings.  This isotropic model we
have investigated previously.\cite{R7}  In the present paper we
extend these studies to the anisotropic case $\eta>1$.

In the model of Eq.(\ref{eHjj}) spatial variations and
fluctuations in the internal magnetic field are ignored, with ${\bf\nabla}
\times {\bf A}=B{\bf\hat z}$ a uniform constant.  This 
should be valid provided  $B$ is so large that 
the magnetic fields associated with each
vortex line strongly overlap, i.e.
\begin{equation}
a_{\rm v}=\sqrt{\phi_0/B}\ll \lambda_\perp \enspace ,
\label{eav}
\end{equation}
where $a_{\rm v}$ is the spacing between vortex lines.  However
$B$ should still be small enough that
$a_{\rm v}\gg\xi_\perp$ (so details of the vortex cores are not
important).  The anisotropy must also be such that
\begin{equation}
d<\lambda_\perp^2/\lambda_z\enspace ,\qquad{\rm or\: 
equivalently,}\qquad 
\eta<\lambda_\perp/\xi_ \perp\enspace ,
\label{emagcoup}
\end{equation}
so that the Josephson coupling between
the planes dominates over the magnetic coupling.\cite{R8,R9}
Formally, our model corresponds to the limit of
$\lambda_\perp/\xi_\perp\to \infty$, keeping $J_\perp$
constant.
For our simulations, we take the $A_{ij}$ as fixed
constants, chosen to give a particular fractional density,
\begin{equation}
    f\equiv B\xi_\perp^2/\phi_0\enspace ,
\label{ef}
\end{equation}
of vortex lines penetrating the $xy$ plane.

Using the model of Eq.(\ref{eHjj}), which is in terms of the
phase angles $\theta_i$, we will also study vortex line
fluctuations.  To locate a vortex line, we compute the phase
angle difference $[\theta_i-\theta_j-A_{ij}]$ across each bond,
restricting this angle to the interval $(-\pi,\pi]$.  The
circulation of these angle differences around any plaquette $\alpha$
must then be $2\pi(n_\alpha -f_\alpha)$, where $f_\alpha = 0$ or $f$
depending on the orientation of the plaquette, and a non-zero integer
value of $n_\alpha$ indicates the presence of a vortex line
piercing the plaquette.  Computing the vorticity of each plaquette
in this fashion, we can then trace out the paths of the vortex lines.

To model a particular material, we would like 
to map out the phase diagram as a function of  $T$ and
magnetic field $B$, for a fixed value of anisotropy $\eta$.
However, due to commensurability difficulties
between the triangular vortex lattice preferred in a continuum and the 
discrete sites permitted by our numerical grid,
different vortex line densities would form lattice structures 
of differing symmetry
in the ground state.  Since we are computationally limited
to a fairly coarse grid, this would make direct comparison of systems
with different $B$ difficult.  We therefore choose to
map out the phase diagram as a function of $T$ and anisotropy $\eta$,
for fixed $B$.  We can see however,  using dimensional arguments, 
that increasing $\eta$ at fixed $B$, is similar to increasing $B$ at fixed $\eta$.
If we measure any
transition temperature $T_{\rm c}$ in units of $J_\perp$, then the dimensionless
$T_{\rm c}/J_\perp$ can only depend on the other dimensionless parameters 
of the Hamiltonian (\ref{eHjj}), 
the anisotropy $\eta=\lambda_zd/\lambda_\perp\xi_\perp$,
and the vortex line density $f=B\xi_\perp^2/\phi_0=(\xi_\perp/a_{\rm 
v})^2$.
Since our London approximation ignores details of the vortex cores, 
if we consider the continuum limit of our model, $a_{\rm v}\gg\xi_\perp$, we
expect that $T_{\rm c}/J_\perp$ should be at most weakly 
dependent\cite{R10}
on the vortex core radius $\xi_\perp$.
The only combination of $\eta$ and $f$ that is independent of 
$\xi_\perp$ is $\eta^2f$.
Thus, the dominant dependence of $T_{\rm c}/J_\perp$ on $\eta$ and $f$ can
only be through some function of $\eta^2f\sim\eta^2B$.

We can further argue how transition temperatures should depend on the 
quantity $\eta^2f$.  In the limit of extreme anisotropy, $\eta\to\infty$,
we have completely decoupled planes, and the transition temperature
should be independent of $\eta$; thus we expect $T_{\rm c}\sim J_\perp$.
In the limit of a nearly isotropic system, $\eta\sim 1$, we expect
that $T_{\rm c}$ should be independent of the spacing between planes $d$;
thus we expect $T_{\rm c}\sim J_\perp/\eta \sqrt 
f=(\phi_0^2/16\pi^3\lambda_\perp^2)(\lambda_\perp/\lambda_z)(\phi_0/B)^{1/2}$.
These are in fact the predictions for the melting temperature based
on Lindemann criterion calculations.\cite{R1,R3,R4}

The cross-over from small to large $\eta$, where the discreteness
of the layering along ${\bf\hat z}$ becomes important and one 
approaches the two dimensional limit, can be estimated
by the criterion 
$\eta_{\rm cr}^2f\simeq 1$, or, using 
$f=B\xi_\perp^2/\phi_0=(\xi_\perp/a_{\rm v})^2$, as
\begin{equation}
\eta_{\rm cr}=a_{\rm v}/\xi_\perp
\label{eetacr}
\end{equation}
Using an effective elastic medium approximation
to describe vortex line fluctuations in the line lattice, one can 
show\cite{R1,R8} that 
for $\eta<\eta_{\rm cr}$, the dominant wavenumber 
$q_z$ of fluctuations at melting 
satisfies the condition $d<\pi/q_z$, and hence the layering of the 
material is averaged over.  For $\eta>\eta_{\rm cr}$ however, the dominant
wavenumber is at $d=\pi/q_z$, and layering is important.  Some theoretical
models\cite{R8,R11,R12}  have predicted that $\eta_{\rm cr}$ 
(or equivalently $B_{\rm cr}=\phi_0\lambda_\perp^2/\lambda_z^2d^2$)
will mark a dramatic change in 
behavior, reflecting a three dimensional to two dimensional cross-over.
Looking for any such cross-over behavior at $\eta_{\rm cr}$ will be one
of the goals of this work.

Our simulations consist of standard Metropolis Monte Carlo simulations of
the Hamiltonian (\ref{eHjj}), using periodic
boundary conditions in all directions, on grid sizes $L_\perp^2
\times L_z$.  
We use a magnetic field $B$ which yields a fractional density of
vortex lines $f=1/15$.  The ground state vortex lattice, 
shown in Fig.\,\ref{f01}, is
a nearly triangular vortex line lattice with sides of length 
$\sqrt{18}\times\sqrt{18}\times\sqrt{17}$ in units of $\xi_\perp$. 
To map out the $\eta-T$ phase diagram, we have done simulations 
varying $T$ at different fixed values of $\eta$ on lattices of size
$15^3$.  We have also carried out simulations of larger system sizes
for the specific cases of $\eta^2=10<\eta^2_{\rm cr}=1/f=15$, and
$\eta^2=50>\eta_{\rm cr}^2$.
Our runs are typically $10,000$ sweeps through the grid
to equilibrate, followed by 128,000 sweeps to compute averages.  These
simulations are about 9 times longer than in our previous 
work.\cite{R7}
Errors are estimated by a standard data blocking procedure.

\section{Numerical Results}
\label{sresults}
\subsection{Phase Diagram}

To test for superconducting coherence, we
compute the helicity moduli $\Upsilon_\perp(T)$ and $\Upsilon_z(T)$ 
which measure the
stiffness with respect to applying a net gradient (``twist'')
in the phase angle of the wavefunction along directions perpendicular
and parallel to the applied magnetic field.\cite{R7}  The helicity modulus
in direction ${\bf\hat\mu}$ is given by the phase angle correlation

\begin{eqnarray}
  \Upsilon_\mu(T)&=&{1\over L_\perp^2L_z}\left\langle\sum_{\langle 
  ij\rangle}J_{ij}\cos(\theta_i-\theta_j-A_{ij})({\bf\hat e}_{ij}\cdot
  {\bf\hat\mu})^2\right\rangle\nonumber\\ &\,&\qquad\qquad\qquad
   -{1\over T L_\perp^2L_z}\left\langle
  \left[\sum_{\langle ij\rangle}J_{ij}\sin(\theta_i-\theta_j-A_{ij})
  ({\bf\hat e}_{ij}\cdot{\bf\hat\mu})\right]^2\right\rangle,
\label{ehelicity}
\end{eqnarray}
where ${\bf\hat e}_{ij}$ is the unit vector from site $i$ to $j$.
When $\Upsilon_\mu$ is positive, the system can carry a 
supercurrent, and so possesses superconducting coherence 
in direction ${\bf\hat\mu}$.  When $\Upsilon_\mu$ vanishes, superconducting
coherence is lost.

To determine the vortex line lattice melting temperature, we compute
the density--density correlation function of vortices within the
same plane,
\begin{equation}
  S({\bf k}_\perp)={1\over L_z}\sum_{{\bf r}_\perp,{\bf 
  r}^\prime_\perp,z}
  {\rm e}^{i{\bf k}_\perp\cdot ({\bf r}_\perp-{\bf r}_\perp^\prime)}
  \langle n_z({\bf r}_\perp,z)n_z({\bf 
  r}^\prime_\perp,z)\rangle\enspace ,
\label{eSk}
\end{equation}
where $n_z({\bf r}_\perp,z)$ is the vorticity at site ${\bf r}_\perp$
in the $xy$ plane at height $z$ (henceforth, we will refer to the vortices
in the $xy$ planes as the ``pancake'' vortices).  
Below melting, we expect to see a periodic
array of sharp Bragg peaks in the ${\bf k}_\perp$ plane.  
Above melting, we expect to see the broad circular rings 
characteristic of a liquid.

We also measure the specific heat per site of the system, $C$, using the usual
energy fluctuation formula.  A peak in $C$ locates
the temperature at which, upon cooling, there is a dramatic freezing
out of thermal fluctuations and the system looses the bulk of
its entropy.  We will take the location
of a high temperature peak (above any phase transitions) in $C$
as indicating the cross-over temperature
$T_{\rm c2}$ where the superconducting wavefunction develops
on small length scales, vortex lines become well defined objects,
and one has the onset of strong diamagnetism.

In Fig.\,\ref{f02} we show our results for $\Upsilon_\perp$ and $\Upsilon_z$
for the case $\eta^2=10$.  We see that
$\Upsilon_\perp$ vanishes at a $T_{\rm c\perp}$ significantly lower than
the $T_{{\rm c}z}$ where $\Upsilon_z$ vanishes. We show data for heating and
cooling, for three different grid sizes, $15^3$, $30^3$, and
$15^2\times 120$.  
Comparing heating and cooling, we see no appreciable hysteresis for
$\Upsilon_z$.  Hysteresis in $\Upsilon_\perp$ appears only for the
$30^3$ system, where we failed to cool back into a lattice.
There are no obvious shifts in $T_{{\rm c}z}$ or $T_{\rm c\perp}$
due to finite size effects as $L_\perp$ and $L_z$ are varied. 
We determine our estimates $T_{\rm c\perp}/J_\perp\simeq 0.36$ 
and $T_{{\rm c}z}/J_\perp\simeq 0.58$ by visually extrapolating
the curves to zero from the inflection point that marks the onset 
of the high temperature tails.  We have found that the 
size of these tails tends to decrease with increasing 
simulation time, as well as with system size.

It is important to note that the finite $T_{\rm c\perp}$ in our
model is strictly an artifact of the discretizing grid, which
acts like an effective periodic pinning potential for
the vortex lines. In a continuum model, one would 
find $\Upsilon_\perp=0$ at {\it all} temperatures,\cite{R13} as the vortex line
lattice is free to slide as a whole, giving ``flux flow resistance.''
A discretizing grid removes this translational symmetry, resulting
in a commensurately pinned vortex line lattice at low temperatures,
with $\Upsilon_\perp >0$.  For a high density of vortex lines,
it is likely that the vortex lattice remains commensurately pinned until
it melts.  In such a case one expects $T_{\rm m}=T_{\rm c\perp}$.
However recent simulations,\cite{R14,R15,R16} 
with a more dilute vortex line density
than studied here, have claimed evidence for a depinning $T_{\rm 
c\perp}$ which is lower than $T_{\rm m}$, with the intermediate
phase a floating vortex line lattice.
It is thus important to determine the
melting $T_{\rm m}$ of our vortex lattice independently from
our measurement of $\Upsilon_\perp$.

In Fig.\,\ref{f03} we show intensity plots at various temperatures
of $S({\bf k}_\perp)$
in the ${\bf k}_\perp$ plane, for the $30^3$ system upon heating.
Looking at when the Bragg peaks disappear, we estimate the
melting temperature to be $T/J_\perp\simeq 0.43$, somewhat higher
than $T_{\rm c\perp}/J_\perp \simeq 0.36$.
To try to quantify the location of the melting transition, we now look
at the heights of the Bragg peaks at the reciprocal lattice vectors.
We denote by $\{{\bf K}_1\}$ the six, almost equal, smallest
non-zero reciprocal lattice vectors.  Let $\{{\bf K}_1^\prime\}$
be the six vectors obtained by reflecting the $\{{\bf K}_1\}$ through
the  ${\bf\hat x}$ axis.  Since the vortex line lattice breaks 
this reflection symmetry of the square discretizing grid, we will
have $S({\bf K}_1)>S({\bf K}_1^\prime)$ for the lattice phase.
However, once the lattice has melted, the reflection symmetry
of the grid should be restored.  We can therefore define as an
order parameter of the melting transition $\Delta S({\bf K}_1)\equiv
S({\bf K}_1)-S({\bf K}_1^\prime)$.  Normalizing by $S_0\equiv S({\bf 
K}=0)$, and averaging over the six $\{{\bf K}_1\}$,
we plot in Fig.\,\ref{f04} $\Delta S({\bf K}_1)/S_0$
versus $T$, for the three system sizes, $15^3$, $30^3$, and 
$15^2\times 120$.  $\Delta S({\bf K}_1)/S_0$
decreases linearly over a large intermediate range of $T$.
From the $30^3$ system we estimate $T_{\rm m}/J_\perp\simeq 0.44$. 
Note that there is a greater finite
size effect and more hysteresis for $\Delta S({\bf K}_1)/S_0$
than there is in $\Upsilon_{\perp,z}$. The estimate for $T_{\rm m}$
tends to decrease as $L_\perp$ increases.  
Our result $T_{\rm c\perp}<T_{\rm m}$ suggests the presence
of a floating vortex  line lattice.  However it remains possible
that $T_{\rm c\perp}$ and $T_{\rm m}$ will merge as the system
size increases,
due either to the finite size dependence observed in $T_{\rm m}$,
or to the possibility that $T_{\rm c\perp}$ actually lies farther out
in the high temperature tail of $\Upsilon_\perp$ than we have
estimated.

Finally, in Fig.\,\ref{f05} we show the specific heat $C$.  The 
high temperature peak in $C$ at $T/J_\perp\simeq 1.0$
we identify with the cross-over $T_{\rm c2}$, which is thus
seen to lie well above $T_{\rm c\perp}$, $T_{\rm m}$, and $T_{{\rm c}z}$.
A suggestion of a smaller peak is seen at the lower temperature
$T_{\rm c\perp}$.

In Fig.\,\ref{f06} we show $\Upsilon_\perp$ and $\Upsilon_z$ for
the case $\eta^2=50$, for system sizes $15^3$ and $30^3$. 
Here the data has considerably more scatter than in Fig.\,\ref{f02}
(in general, we found it increasingly difficult to achieve good
equilibration as $\eta$ increased).  Nevertheless, there again
appears to be two distinct transitions, with $T_{\rm c\perp}/J_\perp\simeq 
0.19<T_{{\rm c}z}/J_\perp\simeq 0.24$.
Intensity plots of the structure function 
$S({\bf k}_\perp)$ are shown in Fig.\,\ref{f07}, and
the peak height differences $\Delta S({\bf K}_1)/S_0$ in Fig.\,\ref{f08}.
These suggest a melting $T_{\rm m}/J_\perp\simeq 0.21$.
In Fig.\,\ref{f09} we show the specific heat $C$.  The
high temperature peak at $T/J_\perp\simeq 1.0$ is again associated 
with $T_{\rm c2}$.  However, comparing with Fig.\,\ref{f05}, there
is now a more clearly defined smaller peak at  $T_{\rm 
c\perp}$.

Carrying out simulations at other values of $\eta$
on a $15^3$ grid, we show in Fig.\,\ref{f10} the resulting phase
diagram in the $\eta-T$ plane. The $T_{{\rm c}z}$ line denotes
the loss of phase coherence parallel to the applied magnetic
field, as measured by the vanishing of $\Upsilon_z$.
The $T_{\rm m}$ line denotes the melting of the vortex line
lattice, as measured by the vanishing of $\Delta S({\bf K}_1)/S_0$.
The $T_{\rm c\perp}$ line denotes the depinning of the vortex
line lattice from the discretizing grid, as measured by the
vanishing of $\Upsilon_{\perp}$.
We see that $T_{\rm m}$ coincides with
$T_{\rm c\perp}$ in the $\eta\sim 1$ and $\eta\gg\eta_{\rm cr}$ limits,
but is somewhat greater than $T_{\rm c\perp}$ in the vicinity of the 
cross-over anisotropy $\eta_{cr}=a_{\rm v}/\xi_\perp=\sqrt{15}$.
Between $T_{\rm m}$ and $T_{{\rm c}z}$ we have
a vortex line liquid which retains superconducting coherence in the
direction parallel to the applied magnetic field.
The dashed line $T_{\rm c2}$ locates the high temperature
peak of the specific heat, and
marks the cross-over from the the vortex line liquid to the normal 
metal.  
The dotted lines labeled $\xi_{\rm c}=n$ will be explained at the end of the
following section.

Thus,
for $T_{\rm c2}<T$ we have the resistive normal metal with weak diamagnetism.
For $T_{{\rm c}z}<T<T_{\rm c2}$ we have a vortex line liquid, with 
strong diamagnetism but still with resistive
behavior in all directions.  For $T_{\rm m}<T<T_{{\rm c}z}$ we have a vortex
line liquid with superconducting coherence parallel to ${\bf B}$.  For
$T<T_{\rm  m}$ we have an Abrikosov vortex line lattice.  For
$T<T_{\rm c\perp}$ the vortex line lattice is pinned.

If we fit the lowest five data points
(those for $\eta\le\eta_{cr}$) to a power law, we find $T_{\rm c\perp}\sim
\eta^{-0.88\pm 0.09}$, $T_{{\rm c}z}\sim \eta^{-0.98\pm 0.05}$,  and 
$T_{\rm m}\sim \eta^{-0.66 \pm 0.07}$.  The results for $T_{\rm 
c\perp}$ and $T_{{\rm c}z}$ are in good agreement with our dimensional
argument that characteristic temperatures at small
$\eta$ should scale as $T\sim \eta^{-1}$.  
The agreement of $T_{\rm m}$ with this form is much poorer.
Whether this reflects the inclusion of too large values of $\eta$ in
the fit, or whether it reflects a poor determination of $T_{\rm m}$
due to finite size effects or incomplete equilibration, remains
unclear.
At large $\eta$, all three lines approach
the constant value $T_{\rm c}^{2\rm D}$, which we have found 
from independent simulations 
to be the melting temperature for an isolated two dimensional 
plane.\cite{R17}

We see that $T_{\rm c2}$
for $\eta>\eta_{cr}$ becomes independent of $\eta$, and is located
at the same temperature as the specific heat peak in
the  ordinary ($B=0$) 2D XY model\cite{R18} (which lies about $10\%$ above
the 2D XY Kosterlitz-Thouless transition at $T_{\rm KT}/J_\perp\simeq 0.9$).  
Thus, at these high temperatures, $\eta_{\rm cr}$ does indeed mark the
dimensional cross-over where our three dimensional system
is behaving as effectively decoupled 2D layers; the cross-over $T_{\rm c2}$
is due to the proliferation of vortex-antivortex pairs within these
decoupled layers.  However, at lower temperatures, we see no dramatic
change in behavior for $T_{\rm c\perp}$, $T_{\rm m}$, and $T_{{\rm c}z}$ 
as $\eta_{cr}$ is crossed.  Layers remain coupled, and
in particular, while $T_{\rm c\perp}$, $T_{\rm m}$, and $T_{{\rm c}z}$ appear
to merge as $\eta$ increases,
we continue to find $T_{\rm c\perp}\le T_{\rm m}\le T_{{\rm c}z}$ 
for all $\eta>\eta_{cr}$ studied.

Note that in the limit of weak anisotropy,
$\eta\to 1$, $T_{{\rm c}z}$ and $T_{\rm c2}$ become close,
as was observed in earlier isotropic simulations.\cite{R7}
However, once the anisotropy $\eta$ increases, $T_{{\rm c}z}$ falls well
below $T_{\rm c2}$.  The transition at $T_{{\rm c}z}$ is thus clearly
distinct from any mean-field-like cross-over phenomena.
We will discuss this point in greater detail in the following section.
Using the analogy between increasing $\eta$ and increasing $B$ as
discussed in Section \ref{smodel}, the increase in the width  of the 
vortex line liquid region in Fig.\,\ref{f10}, as anisotropy increases, 
is in agreement with the general experimental features discussed in the
Introduction.

\subsection{Vortex Line Fluctuations}

We now discuss several measures of the vortex line
fluctuations in our model, in order to try and clarify the nature
of the phenomena at $T_{{\rm c}z}$ and $T_{\rm c2}$.

The first quantity we consider is $\Delta\ell_\mu$, defined as 
the the total number of vortex line segments due to fluctuations
in direction ${\bf\hat\mu}$, normalized by the total
number of field induced
``pancake'' vortices in the $xy$ planes, $fL_\perp^2L_z$.
Note that in computing $\Delta\ell_\mu$, line segments are added
without regard to the sign of their direction; oppositely oriented segments
do {\it not} cancel out.  In Figs.\,\ref{f11}$a$ and \ref{f11}$b$ we show
our results for $\Delta\ell_\perp\equiv {1\over 2}(\Delta\ell_x
+\Delta\ell_y)$ and $\Delta\ell_z$ for the two cases of
$\eta^2=10$ and $\eta^2=50$ respectively.  We see that in 
both cases, $\Delta\ell_z$ is at least two orders of magnitude
smaller than $\Delta\ell_\perp$ in the vicinity of
$T_{{\rm c}z}$ and below.  Thus, only transverse
vortex fluctuations appear to be important at the phase transitions.
Only at the higher cross-over $T_{\rm c2}$, does $\Delta\ell_z$
start to become comparable to $\Delta\ell_\perp$.  This is consistent
with our interpretation of $T_{\rm c2}$ as the temperature at which
vortex-anitvortex pairs start to enter the $xy$ planes.

One possible explanation for the transition at $T_{{\rm c}z}$
has been proposed by Nelson,\cite{R3} in terms of the entanglement of
vortex lines.  If we assume, as in Nelson's picture, that
the transverse fluctuation of a vortex line in the liquid phase
is like that of a random walk, then since $\Delta\ell_\perp$ is the
net transverse fluctuation per pancake vortex, the total
transverse deflection of a line in traveling down the length of the
system will be $u=\sqrt{L_z}\Delta\ell_\perp$.  Geometric 
entanglement\cite{R18.5} of lines should occur
when $u\simeq a_{\rm v}$, or when $\Delta\ell_\perp\simeq a_{\rm v}/
\sqrt{L_z}=1/\sqrt{fL_z}$.  For our system with $f=1/15$, this
criterion gives entanglement at values of $\Delta\ell_\perp=1.0$,
$0.71$, and $0.35$ for thicknesses $L_z=15$, $30$, and $120$ 
respectively.  Noting that $\Delta\ell_\perp$ shows no apparent
dependence on $L_z$ near the transitions, we would conclude
that geometric entanglement takes place noticeably {\it below} 
$T_{{\rm c}z}$ for
systems of thickness $L_z>15$.  It is interesting to note that
in both cases Fig.\,\ref{f11}$a$ and \ref{f11}$b$, $T_{{\rm c}z}$ appears
to coincide with the point where $\Delta\ell_\perp\simeq 1$.
We have similarly observed this to be true at other values of
$\eta$.  However we have no explanation for this coincidence.

The above argument assumed that all of the vortex line
fluctuations consisted of transverse motions of the magnetic field
induced vortex lines.  However, there is additionally the possibility
of forming thermally excited closed vortex rings, which for 
large enough $\eta$ and temperatures
low compared to $T_{\rm c2}$ should tend to lie between two adjacent $xy$ planes.
We now describe our algorithm to trace out the paths of vortex lines,
which will allow us to measure both the  distribution of such closed rings,
as well as the entanglement of the field induced lines.
We start by searching the plaquettes for a 
penetrating vortex line segment.  We then trace
its path into and out of subsequent unit cells of the grid.  Such
a line can belong either to a field induced vortex line, 
or to a closed vortex ring.  Tracing the line, we measure
the net displacement parallel to ${\bf\hat z}$ that is traveled
before the line closes back upon itself.  If we have a closed ring,
this net displacement is zero, and we measure the perimeter of the 
ring $p$.  If we have a field induced line, then because of our 
periodic boundary conditions parallel to ${\bf\hat z}$, this
net displacement must be $mL_z$ with integer $m=1,2,\ldots,fL_\perp^2$.
If $m=1$ the line closes back upon itself upon traversing the length
of the system $L_z$.  For $m>1$, the line
belongs to a group of $m$ lines that are braided with each other.
This is schematically illustrated in Fig.\,\ref{f12}.  
The distribution of values of $m$ is a measure of how geometrically 
entangled the
field induced lines are.  With this procedure, we search
through all plaquettes until all vortex line segments are found, and
classified as belonging to either a ring of perimeter $p$, or
an entangled braid of order $m$.
The only complication in the above algorithm occurs when
two or more vortex lines segments intersect, i.e. go in and out of the same
unit cell of the grid.  In this case we randomly choose which segment 
is connected
to which.  In practice this was achieved as follows.  Once a
line was traced into a unit cell, we searched the remaining five
faces in a random order to see which face the line is leaving through.
Once we find a line leaving, we take it to be the continuation of
the line we are tracing.

In Figs.\,\ref{f13}$a$ and \ref{f13}$b$ we show our results for
the distribution $q(p)$ of the number of closed rings of perimeter
$p$ per unit volume $L_zL_\perp^2$,
for the two cases $\eta^2=10$ and $\eta^2=50$ (we show results for
cooling; no significant hysteresis was observed comparing heating and cooling).  
Plotting the logarithm
of $q(p)$ versus $1/T$, we see approximately straight lines at low 
$T$, indicating thermal activation.  These lines
have a change in slope in the vicinity of the melting $T_{\rm m}$, which is
mild for $\eta^2=10$, but more pronounced for $\eta^2=50$.
At the higher $T_{\rm c2}$ the curves saturate.  We believe this 
is consistent with the interpretation of $T_{\rm c2}$ as the
cross-over temperature at which, upon heating, vorticity explodes
throughout the system, and superconducting order is lost on even
small length scales.  The saturation of $q(p)$ occurs because, above 
$T_{\rm c2}$, the
distribution of ring sizes is governed more by the statistics
of random intersections among the lines, rather than by energetics.
Note that in Ref.\onlinecite{R7}, where only the isotropic case was
studied, we incorrectly associated this 
explosion of vorticity, as indicated by the saturation of $q(p)$,
with the transition at $T_{{\rm c}z}$.  From the phase diagram of
Fig.\,\ref{f10}, we now see that this mistake 
was due to the proximity of $T_{\rm c2}$ and $T_{{\rm c}z}$ which 
occurs only in the
isotropic limit.  For anisotropic systems, $T_{{\rm c}z}$ drops
below $T_{\rm c2}$ and lies in the region where $q(p)$ 
is still governed by thermal activation.

From the data of Figs.\,\ref{f11} and \ref{f13}, we can now
compare how much of the vortex line fluctuations is contained in the
wandering of the field induced vortex lines, versus how much is contained in 
the thermally excited vortex rings.  We show in Figs.\,\ref{f14}$a$
and \ref{f14}$b$ the total length of {\it all} vortex line
fluctuations (per pancake vortex), $\Delta\ell_{\rm tot}\equiv
2\Delta\ell_\perp+\Delta\ell_z$, and the total length contained
in closed vortex rings (per pancake vortex), $\Delta\ell_{\rm ring}
\equiv f^{-1}\sum_p pq(p)$, for the cases $\eta^2=10$ and 
$\eta^2=50$ respectively.  We see that at low temperatures, rings
constitute a negligible fraction of the total vortex line fluctuations.
At the transition $T_{{\rm c}z}$ they are only about $3.5\%$ of
the fluctuations for $\eta^2=10$, and $1.5\%$ for $\eta^2=50$.
It thus seems that as anisotropy increases, the importance of
disconnected thermally excited vortex rings decreases.  It
is worth noting however that ring excitations
which connect to field induced lines,
may still play a role in determining the wandering
of the magnetic field induced lines between planes.
An example is illustrated in Fig.\,\ref{f15}.  Such ``connected'' ring
excitations, which lie between $xy$ planes, are degrees
of freedom distinct from the pancake vortices which lie within
the planes.  In Figs.\,\ref{f16}$a$ and \ref{f16}$b$, we show
snapshot views of vortex line configurations, at various temperatures,
for the cases $\eta^2=10$ and $\eta^2=50$, for a system of size $15^3$.
The bottom row in each figure is a view of all vortex line segments
that lie between a typical pair of adjacent $xy$ planes.
A ``$\sqcup$'' shaped segment in these bottom row views indicates
a connected ring excitation.  We see that they
are present in the system at virtually all temperatures shown,
however it remains unclear how, if at all, they correlate with the
transitions.

We turn now to consider the fluctuations of the magnetic field
induced vortex lines.  As mentioned in connection with our line
tracing algorithm, we can classify each such line as belong to 
an entangled braid of $m$ lines, $m=1,\ldots,fL_\perp^2$ (see
Fig.\,\ref{f12}).  
We denote by $n(m)$ the number of lines which participate in
a braid of order $m$, and $R\equiv n(1)/fL_\perp^2$ is the fraction
of unentangled lines.  In Figs.\,\ref{f17}$a$ and \ref{f17}$b$ we
plot $R$ versus $T$ for the cases $\eta^2=10$ and $\eta^2=50$
respectively.  We see that upon heating, $R=1$ up to $T_{\rm m}$, 
throughout the vortex line lattice phase.  Above $T_{\rm m}$,
$R$ starts to drop, tending to saturate to its high $T$ limit
around $T_{{\rm c}z}$.  Upon cooling, $R$ starts to increase
at $T_{{\rm c}z}$, and in most cases reaches a completely
disentangled configuration with $R=1$ at $T_{\rm m}$.
These features were found at all values of $\eta$ studied.
For the thickest sample of $L_z=120$ at
$\eta^2=10$, however, we cool into a glassy entangled state
with $R\simeq 0.47$ frozen below $T_{\rm c\perp}$, 
as has been seen in previous isotropic
simulations.\cite{R7}  From Figs.\,\ref{f17} it seems clear that the 
transition at $T_{{\rm c}z}$ is related to the braiding
of lines.

To see this another way, we plot in Figs.\,\ref{f18}$a$ and \ref{f18}$b$,
for $\eta^2=10$ and $\eta^2=50$ respectively,
the braid distribution $n(m)$ versus $m$, for several different
temperatures near $T_{{\rm c}z}$.  We use our data for systems
of size $30^3$, which have $60$ lines.  We see that for $T<T_{{\rm 
c}z}$, $n(m)$ is strongly peaked at small $m$, decaying rapidly
as $m$ increases.  However, as $T_{{\rm c}z}$ is approached,
the peak at small $m$ decreases and
the distribution $n(m)$ becomes flat and equal to unity for an
increasingly wide range of intermediate $m$. 
The transition at $T_{{\rm c}z}$ 
therefore seems to be associated with braids involving a macroscopically
large number of lines.  When $n(m)=1$ for {\it all} $m$, it indicates
that a line $i$ which starts out at ${\bf r}_{i\perp}(z=0)$ in the $xy$
plane at $z=0$, is equally likely to match onto the starting position
of any other line $j$, after traveling down the thickness of the system, i.e.
${\bf r}_{i\perp}(z=L_z)={\bf r}_{j\perp}(z=0)$ is equally likely for
any pair $i$ and $j$ (see Fig.\,\ref{f12}).  We may speculate that
precisely this condition is achieved at $T_{{\rm c}z}$ in the limit
of large system sizes, and long simulation times.

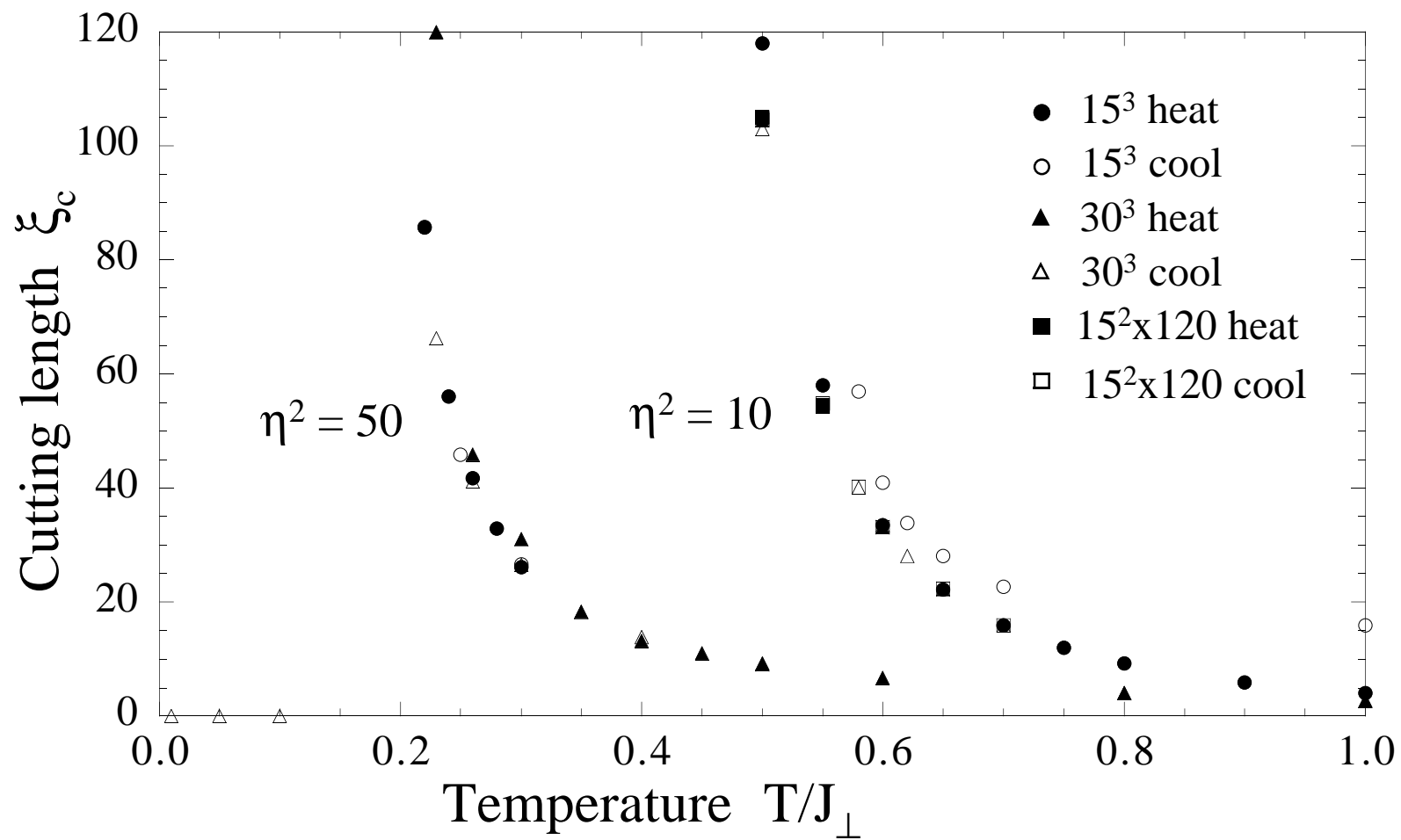
\begin{figure} 
\caption{Cutting length $\xi_{\rm c}$ vs. $T$ for $\eta^2=10$, and $50$,
for various system sizes.}
\label{f19}
\end{figure}

An intriguing question concerning behavior in the vortex
line liquid is how easily lines can cut through each other.
This has important consequences for line diffusion.  If lines cannot
cut, they can be effectively pinned by their mutual 
entanglements.\cite{R3,R19} 
For our system in particular, with periodic boundary
conditions parallel to the magnetic field, the degree of entanglement
can only change due to the cutting and reconnecting of lines.
To investigate this we have computed the average number of vortex 
line intersections, $N_{\rm c}$, present in any instantaneous configuration 
of the system. An intersection is defined when two vortex lines
enter and leave the same unit cell of the grid, and corresponds to
vortex lines with overlapping cores.  Once two lines  
intersect, they are free to cut through each other, or even to
detach and reconnect different ingoing and outgoing segments.
We define the ``cutting length'' $\xi_{\rm c}\equiv fL_\perp^2L_z/N_{\rm c}$
as the average distance (in units of $d$) along ${\bf\hat z}$ between cuts 
of the magnetic field induced vortex lines.  $\xi_{\rm c}$ gives a crude measure
of the average length over which a vortex line remains a well defined
string, or equivalently a measure of the number of planes which remain 
correlated.  In Fig.\,\ref{f19} we plot $\xi_{\rm c}$ versus $T$ for the
two cases $\eta^2=10$ and $\eta^2=50$.  In the phase diagram of 
Fig.\,\ref{f10} we show contours of constant
$\xi_{\rm c}=2$, $4$, $6$, and $10$.  We see that
planes are essentially uncorrelated at temperatures above the 
cross-over $T_{\rm c 2}$.  However correlations grow and get
large as one cools below $T_{\rm c2}$ towards $T_{{\rm c}z}$.
The picture presented by the
contours of $\xi_{\rm c}$ in Fig.\,\ref{f10}, combined with the
behavior of $R$ in Figs.\,\ref{f17} and $n(m)$ in Figs.\,\ref{f18},
is that intersections start to freeze out
below $T_{\rm c2}$, with lines becoming well defined
on longer and longer length scales. This presumably will affect the
time scales on which lines are able to diffuse about, with a
a corresponding signature to be expected in dynamic phenomena.
However the equilibrium degree of 
entanglement, as measured by $R$ and $n(m)$ remains largely unchanged
down to $T_{{\rm c}z}$.  Below $T_{{\rm c}z}$, the behavior of $R$ 
and $n(m)$ shows that the lines start to disentangle, yet
cutting is still frequent
enough to change the degree of entanglement for all temperatures 
down to $T_{\rm m}$.
Below $T_{\rm m}$ the lines remain either in a disentangled
lattice phase, or a metastable state with a frozen degree of entanglement.

\section{Discussion}
\label{sdisc}

We have computed the phase diagram of a fluctuating type II superconductor
in the anisotropy--temperature plane.  Our results are consistent with
general experimental observations, that vortex lattice melting occurs
well below the cross-over $T_{\rm c2}$ associated with the formation of local
superconducting order, and that the width of this region increases
with increasing magnetic field (anisotropy).
As was found in earlier isotropic simulations, we continue
to find as anisotropy increases that there exists a finite region of the 
vortex line liquid which
possesses superconducting behavior parallel to the applied magnetic 
field.  Such a region has not been seen in other simulations\cite{R20}
which have used a higher vortex line density, $f=1/5$, $1/6$, and $1/8$.
The transition at $T_{{\rm c}z}$ from the superconducting line
liquid to the normal line liquid appears to be associated with the
braiding of a macroscopically large number of the field induced
vortex lines, in qualitative agreement with Nelson's picture based on
an analogy with the superfluid transition of two dimensional 
bosons.\cite{R3}  
However, unlike Nelson's picture, $T_{{\rm c}z}$ does not appear to 
decrease with increasing $L_z$.  Such a possibility has been proposed
by Feigel'man and co-workers\cite{R21} in terms of an analogy to 2D bosons with
long range interactions.
Recently, Te\v{s}anovi\'c has proposed\cite{R22} a mechanism for such a 
transition in terms of a vortex loop unbinding analog of the
transition at $B=0$.  Our numerical results indicate that
disconnected thermally induced closed vortex rings
do not appear to be playing any significant role at $T_{{\rm c}z}$,
once the anisotropy has increased enough that $T_{{\rm c}z}$ is
significantly below $T_{\rm c2}$.  However it remains
unclear whether or not  vortex rings between planes, which are
connected to the
field induced lines (see Figs.\,\ref{f15} and \ref{f16}), are
important degrees of freedom.

We have studied behavior as the anisotropy is increased
beyond the ``3D-2D'' cross-over, which simple dimensional
analysis gave as $\eta^2_{\rm cr}f=1$.  We found that this
did indeed mark the cross-over from $\eta$--dependent 
(3D) behavior to $\eta$--independent (2D) behavior, at the high
cross-over temperature $T_{\rm c2}$.  $\eta_{\rm cr}$ also
corresponds to the region where there is the widest relative width 
for the floating
vortex line lattice, $T_{\rm c\perp}<T<T_{\rm m}$.
However we see no qualitative changes in critical behavior as
$\eta_{\rm cr}$ is crossed.  As $\eta$ increases above
$\eta_{\rm cr}$, $T_{\rm c\perp}$, $T_{\rm m}$, and $T_{{\rm 
c}z}$ all approach each other, but
we continue to find $T_{\rm c\perp}\le T_{\rm m}\le T_{{\rm c}z}$.

Glazman and Koshelev,\cite{R8} by considering the effect of elastic distortions
of the vortex line lattice on interplanar phase fluctuations, have
argued that $T_{{\rm c}z}$ should decrease below $T_{\rm m}$,
with a dependence $T_{{\rm c}z}\sim\eta^{-1}$,
when the anisotropy increases above a value $\sim 10\eta_{\rm cr}$.
Frey et al.\cite{R12} have argued that when $\eta\gg\eta_{\rm cr}$,
the proliferation of vortex lattice defects will create a 
``supersolid'' phase, leading to $T_{{\rm c}z}\sim 1/\ln\eta$, 
which again falls below $T_{\rm m}$ for large enough $\eta$.
Similar results were earlier proposed by Feigel'man et al.\cite{R11}
We were unable to equilibrate our system at such high anisotropies
so as to more thoroughly check these predictions
(very large values of $\eta$ require relatively large values of 
$L_\perp$, so that the {\it total} interplanar coupling energy remains
large compared to $T$).
Recently, simulations\cite{R15} by Nguyen et al., of a vortex line model with a 
finite value of $\lambda_\perp<a_{\rm v}$, find evidence for $T_{{\rm 
c}z}<T_{\rm m}$ for moderate anisotropies.  
They find $T_{{\rm c}z}\sim\eta^{-2}$,
scaling with the coupling between planes, in contrast to
the above two theoretical predictions. They attribute the effect
as due to the proliferation of closed vortex rings lying between
planes, in contrast with our own findings that such rings do
not exist in significant numbers.  The point where
their $T_{{\rm c}z}$ crosses below $T_{\rm m}$ occurs at the
value of anisotropy where magnetic coupling between planes
starts to dominate over Josephson coupling, 
$\eta>\lambda_\perp/\xi_\perp$.  Their results thus 
lie outside the range of validity of our infinite 
$\lambda_\perp$ model (see Eq.(\ref{emagcoup})).  
Simulations by \v{S}\'{a}\v{s}ik and
Stroud,\cite{R23} on an anisotropic model in the ``Lowest Landau Level''
approximation (also a $\lambda_\perp\to\infty$ approximation),
always find $T_{{\rm c}z}= T_{\rm m}$ for all anisotropies
studied.  Thus the possibility that parallel coherence can vanish
at a {\it lower} temperature than melting, and if so the nature
of the mechanism responsible, remain issues yet to be clarified
numerically.

Finally, we have shown that
the effective length over which vortex lines can be considered
well defined connected objects, as measured by the distance
between intersections $\xi_{\rm c}$, steadily
increases once one cools below $T_{\rm c2}$ into the vortex line
liquid.  This is consistent with the analysis of ``non-local'' 
conductivity in flux transformer experiments, which indicate that
correlations parallel to $B$ start to grow right from the onset
of strong diamagnetism.\cite{R24}
Nevertheless, we find that vortex line cutting remains sufficiently
easy over most of the vortex line liquid region.  This is
illustrated by the absence of any hysteresis in the 
measurement of our entanglement parameter $R$, for most of the temperature
range $T_{\rm m}<T$ (recall that for our periodic boundary
conditions, $R$ can only change value due to the cutting and 
reconnection of lines).  Only for $\eta^2=10$ and our thickest
$L_z=120$, did we find freezing into a non-equilibrium
state with finite entanglement, that sets in near $T_{\rm m}$, and
saturates below $T_{\rm c\perp}$.  In earlier isotropic 
simulations\cite{R7}, this freezing out of equilibrium was
found to occur at a higher temperature, a little below $T_{{\rm c}z}$.

\section*{Acknowledgment}
We would like to thank A. E. Koshelev, C. Ciordas-Ciurdariu,
and Y.-H. Li, for valuable discussions and assistance.
This work has been supported by the U.S. Department of Energy under
grant DE-FG02-89ER14017.

\newpage

\begin{figure}
\caption{Ground state locations of vortex
lines in the $xy$ plane for line density $f=1/15$ on a cubic grid.}
\label{f01}
\end{figure}

\begin{figure} 
\caption{Helicity moduli $\Upsilon_\perp$ and $\Upsilon_z$
versus temperature $T$ for anisotropy
$\eta^2=10$ and vortex line density $f=1/15$.  Heating and
cooling data for three different system sizes are shown, along with
representative errors.}
\label{f02}
\end{figure}

\begin{figure} 
\caption{Structure function $S({\bf k}_\perp)$ for $\eta^2=10$ and
$f=1/15$ for system size $30^3$, upon heating.  
The cross-over from Bragg peaks to
liquid like rings occurs at $T_{\rm m}/J_\perp\simeq 0.43$}
\label{f03}
\end{figure}

\begin{figure} 
\caption{Bragg peak heights $\Delta S({\bf K}_1)/S_0$ for $\eta^2=10$,
$f=1/15$, and different system sizes.}
\label{f04}
\end{figure}

\begin{figure} 
\caption{Specific heat $C$ versus $T$ for $\eta^2=10$, $f=1/15$, and
various system sizes. The high temperature peak locates the 
cross-over $T_{\rm c2}$.}
\label{f05}
\end{figure}

\begin{figure} 
\caption{Helicity moduli $\Upsilon_\perp$ and $\Upsilon_z$
versus temperature $T$ for anisotropy
$\eta^2=50$ and vortex line density $f=1/15$.  Heating and
cooling for two different system sizes are shown, along with
representative error bars.}
\label{f06}
\end{figure}

\begin{figure} 
\caption{Structure function $S({\bf k}_\perp)$ for $\eta^2=50$ and
$f=1/15$ for system size $30^3$, upon heating.  
The cross-over from Bragg peaks to
liquid like rings occurs at $T_{\rm m}/J_\perp\simeq 0.21$}
\label{f07}
\end{figure}

\begin{figure} 
\caption{Bragg peak heights $\Delta S({\bf K}_1)/S_0$ for $\eta^2=50$,
$f=1/15$, and different system sizes.}
\label{f08}
\end{figure}

\begin{figure} 
\caption{Specific heat $C$ versus $T$ for $\eta^2=50$, $f=1/15$, and
various system sizes. The high temperature peak locates the cross-over 
$T_{\rm c2}$.  A lower temperature peak corresponds to $T_{\rm 
c\perp}$}
\label{f09}
\end{figure}

\begin{figure} 
\caption{Phase diagram in the
anisotropy--temperature plane for vortex line density
$f=1/15$.  $\xi_{\rm c}$  is measured in units 
of $d$. $T_{\rm c2}$ locates the peak in specific heat.}
\label{f10}
\end{figure}

\begin{figure} 
\caption{Excess vortex line length due to fluctuations in transverse,
$\Delta\ell_\perp$, and parallel, $\Delta\ell_z$, directions.  $a$) is
for $\eta^2=10$, $b$) is for $\eta^2=50$.  Dashed lines are guides to 
the eye only.}
\label{f11}
\end{figure}

\begin{figure} 
\caption{Schematic example of the possible reconnections of field
induced vortex lines, under application of the periodic boundary
condition in the ${\bf\hat z}$ direction.  Solid, dashed, and dotted
lines are used to distinguish the different lines within a 
particular braid.}
\label{f12}
\end{figure}

\begin{figure} 
\caption{Number of closed vortex rings $q(p)$ of perimeter $p$, per
unit volume.  $a$) is
for $\eta^2=10$, $b$) is for $\eta^2=50$.}
\label{f13}
\end{figure}

\begin{figure} 
\caption{Total length of all vortex line fluctuations, $\Delta\ell_{\rm 
tot}$, and total length of lines in closed vortex rings, 
$\Delta\ell_{\rm ring}$, per number of pancake vortices.  $a$) is
for $\eta^2=10$, $b$) is for $\eta^2=50$.}
\label{f14}
\end{figure}

\begin{figure} 
\caption{Schematic example of how ``connected''  vortex rings between
planes contribute to the wandering of field induced lines
between planes.}
\label{f15}
\end{figure}

\begin{figure} 
\caption{Snapshot views of vortex line configurations at various
temperatures for $a$) $\eta^2=10$, and $b$) $\eta^2=50$, for $15^3$
size system.  For each case we show perspective views from the side
(top row), 
and looking straight down along the applied field (middle row).  Solid dots
indicate a point of intersection between two vortex line segments.
The bottom row is a view of all vortex line segments that lie
between a typical pair of adjacent $xy$ planes.}
\label{f16}
\end{figure}

\begin{figure} 
\caption{Fraction of unentangled lines $R$, vs. $T$, for various
system sizes. $a$) is
for $\eta^2=10$, $b$) is for $\eta^2=50$.}
\label{f17}
\end{figure}

\begin{figure} 
\caption{Braid distribution $n(m)$ vs. $m$ for several
temperatures near $T_{{\rm c}z}$, for system size $30^3$.  $a$) is
for $\eta^2=10$, $b$) is for $\eta^2=50$.}
\label{f18}
\end{figure}

\begin{references}
\bibitem{R17} For a 2D vortex lattice in a continuum, the melting
  temperature is at $T_{\rm c}^{\rm 2D}=0.044 J_\perp$.  The rather
  higher value found here is due to the effects of the discretizing
  grid.  See M. Franz and S. Teitel in Ref.\,\onlinecite{R16}.
\bibitem{R18.5} We refer here to ``geometric'' entanglement as meaning the
   geometrical twisting of lines around each other.  This is in 
   contrast to other, more specific, definitions of entanglement which
   rigorously relate to the existence of superfluidity in the analog system of
   2D bosons (see for example Ref.\onlinecite{R12}).
\end{references}
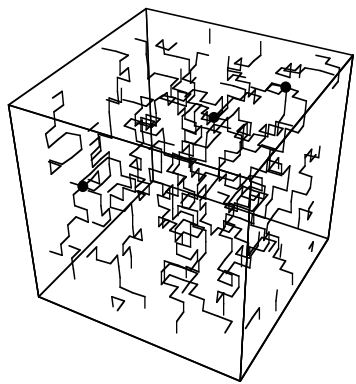
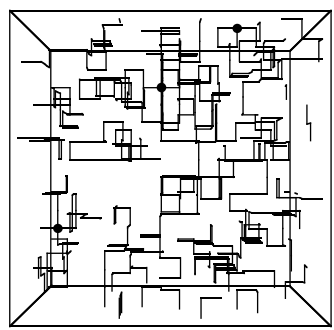
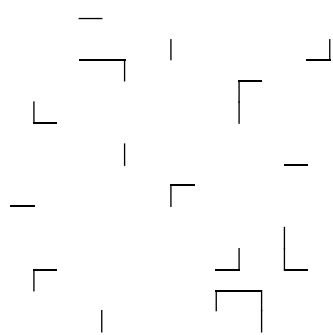
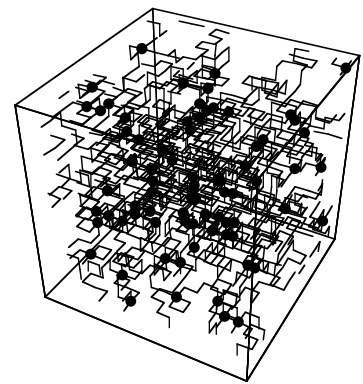
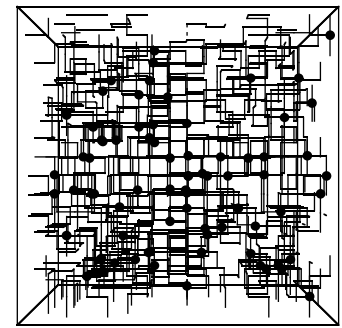
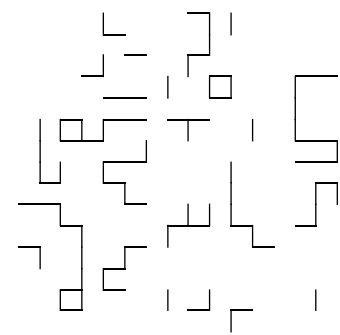
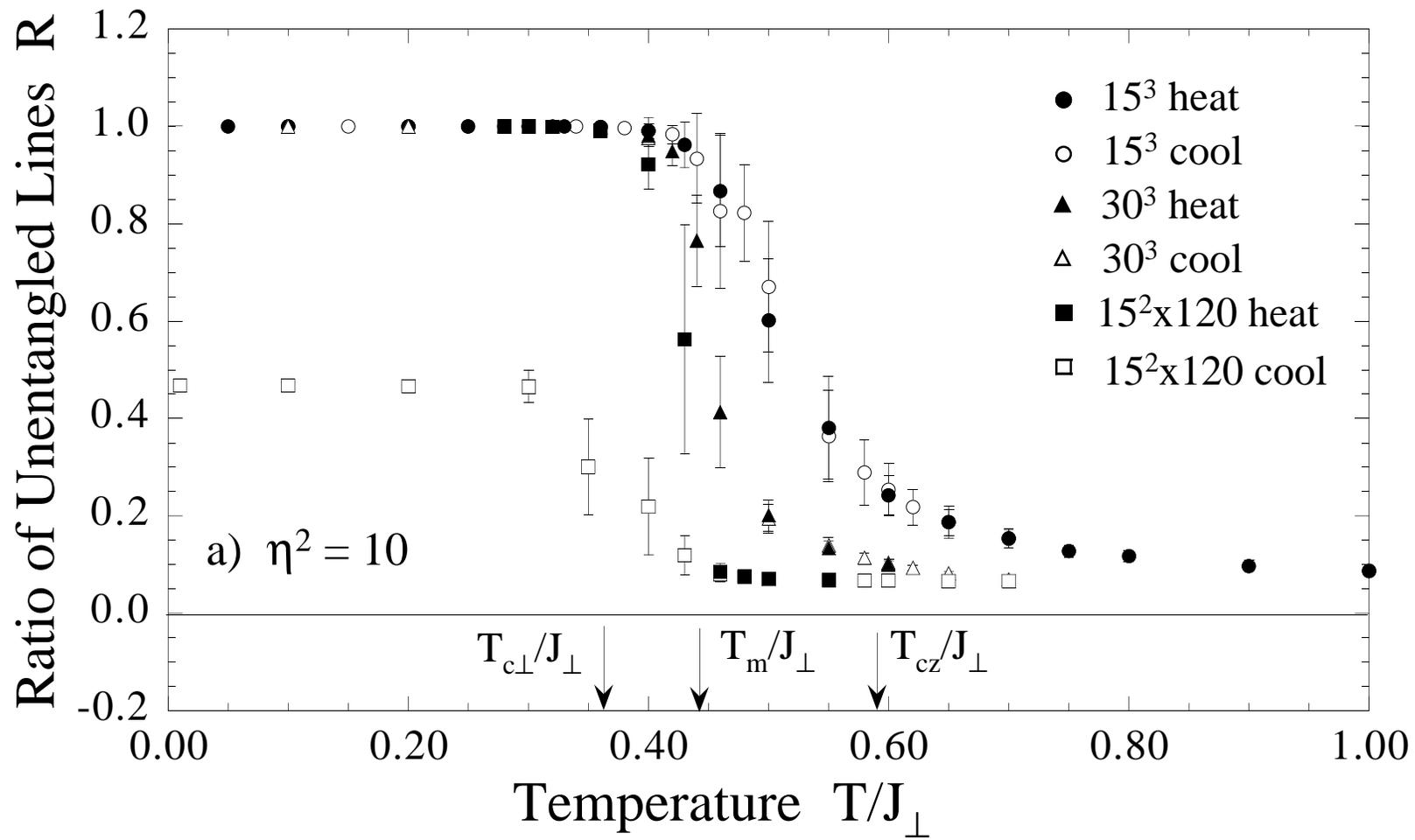
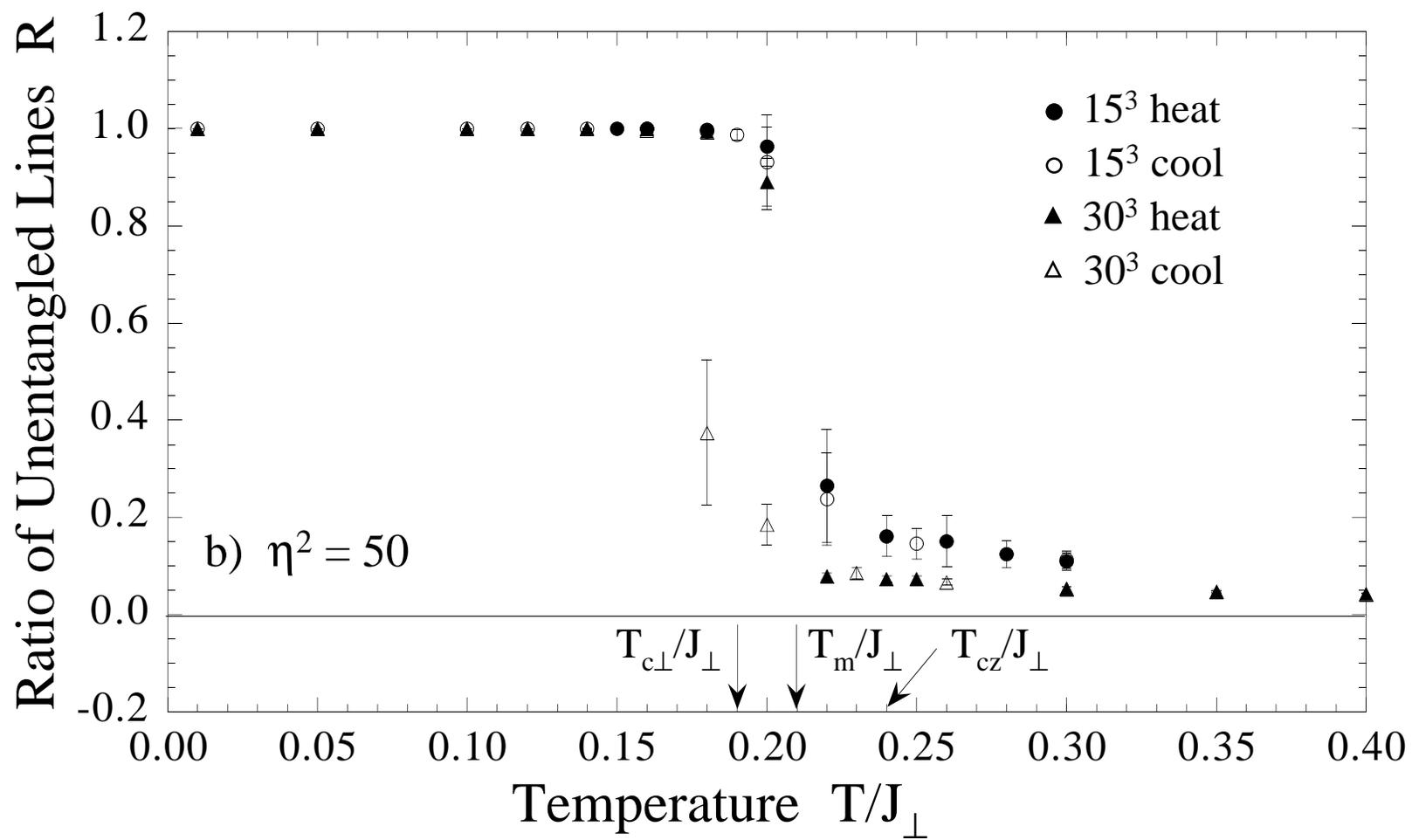
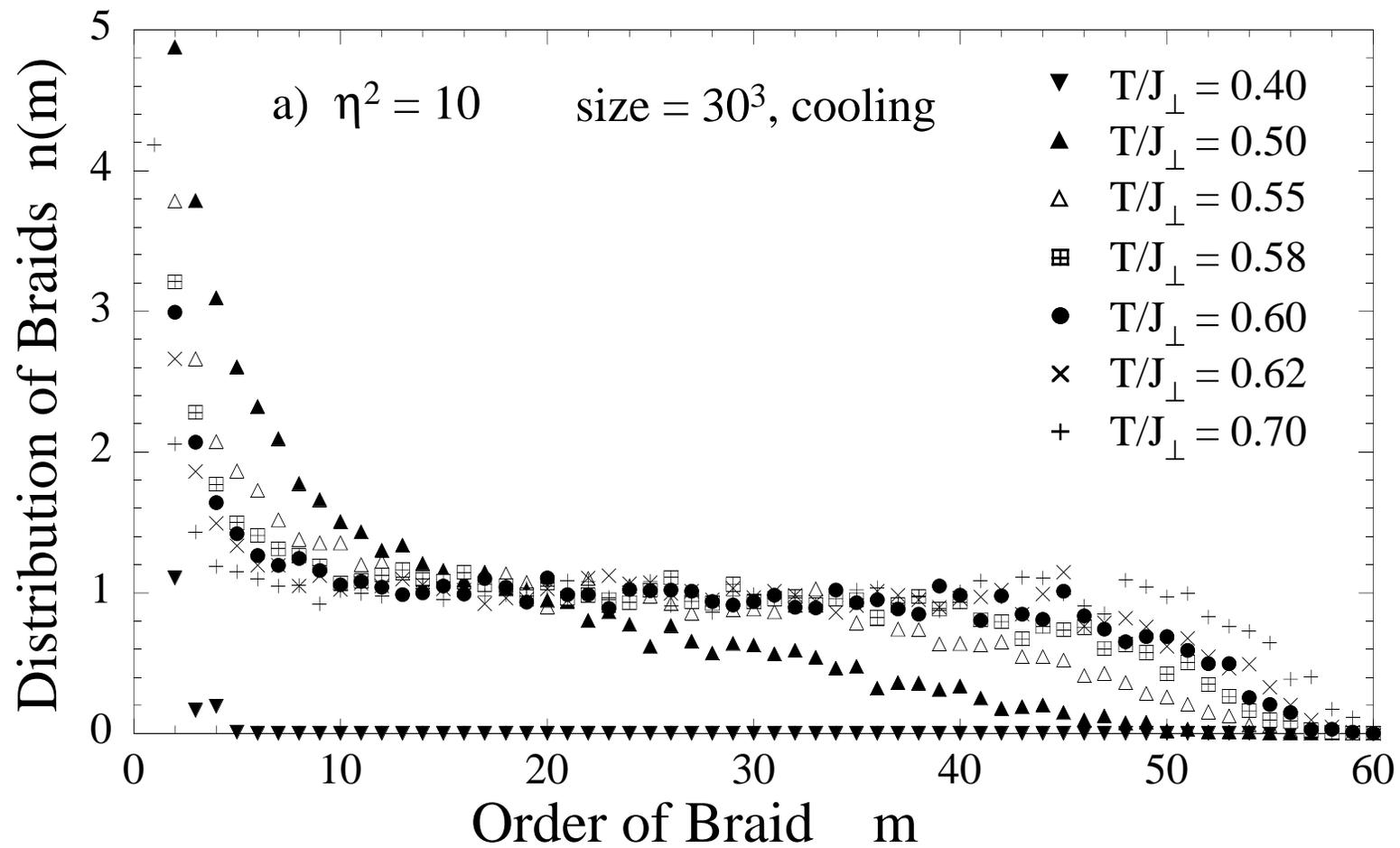
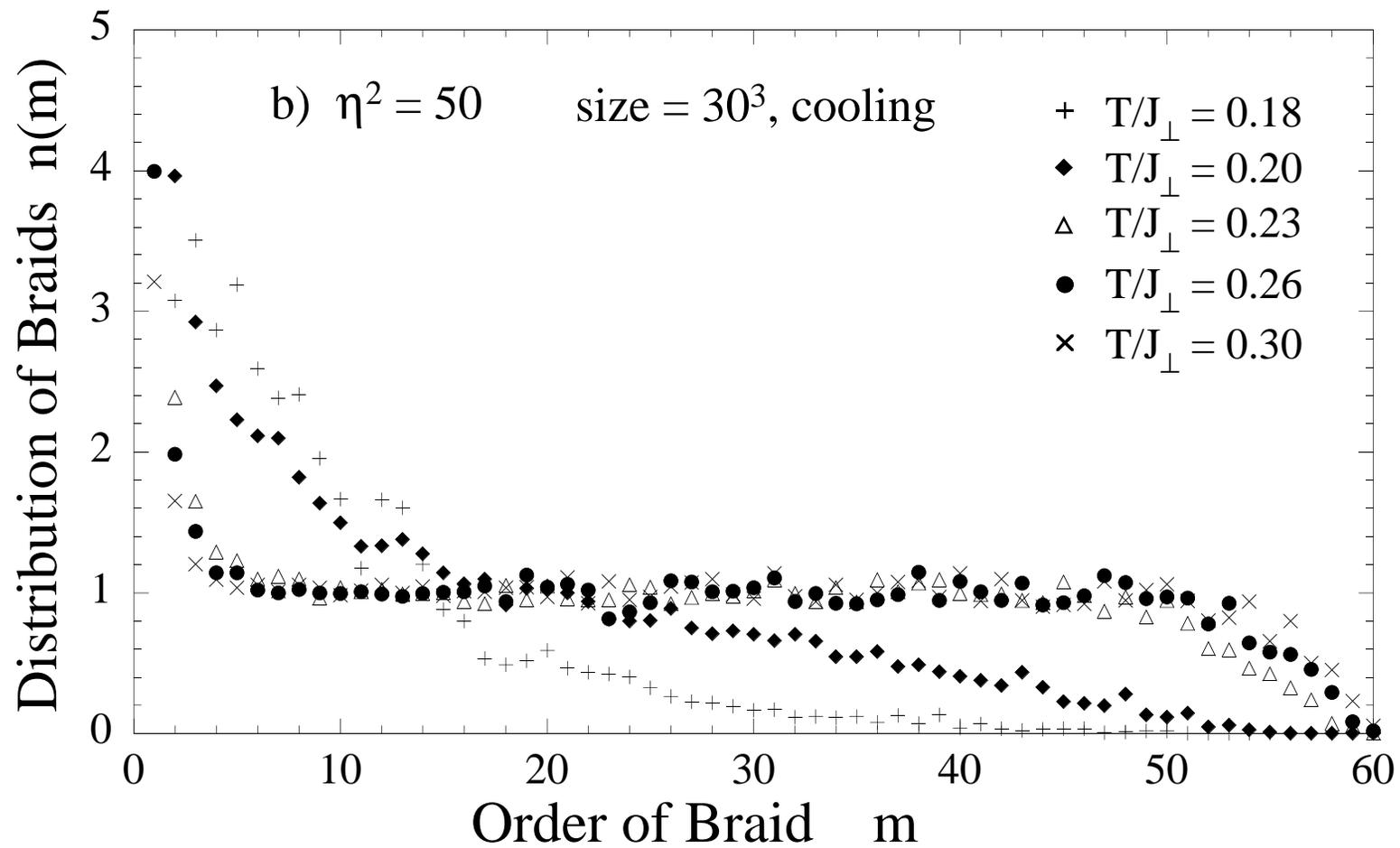
\end{document}